\documentclass[12pt,preprint]{aastex}

\shorttitle{Chromospheric Evaporation Observed with IRIS and EIS}
\shortauthors{Li et al.}

\begin{document}

\title{Chromospheric Evaporation in an X1.0 Flare on 2014 March 29 Observed with IRIS and EIS}
\author{Y. Li$^{1,2}$, M. D. Ding$^{1,2}$, J. Qiu$^{3}$, J. X. Cheng$^{4}$}

\affil{$^1$School of Astronomy and Space Science, Nanjing University, Nanjing 210093, China}
\affil{$^2$Key Laboratory for Modern Astronomy and Astrophysics (Nanjing University), Ministry of Education, Nanjing 210093, China}
\affil{$^3$Department of Physics, Montana State University, Bozeman, MT 59717, USA}
\affil{$^4$Shanghai Astronomical Observatory, Chinese Academy of Sciences, Shanghai 200030, China}
\email{yingli@nju.edu.cn}

\begin{abstract}
Chromospheric evaporation refers to dynamic mass motions in flare loops as a result of rapid energy deposition in the chromosphere. These have been observed as blueshifts in X-ray and extreme-ultraviolet (EUV) spectral lines corresponding to upward motions at a few tens to a few hundreds of km s$^{-1}$. Past spectroscopic observations have also revealed a dominant stationary component, in addition to the blueshifted component, in emission lines formed at high temperatures ($\sim$10 MK). This is contradictory to evaporation models predicting predominant blueshifts in hot lines. The recently launched Interface Region Imaging Spectrograph (IRIS) provides high resolution imaging and spectroscopic observations that focus on the chromosphere and transition region in the UV passband. Using the new IRIS observations, combined with coordinated observations from the EUV Imaging Spectrometer, we study the chromospheric evaporation process from the upper chromosphere to corona during an X1.0 flare on 2014 March 29. We find evident evaporation signatures, characterized by Doppler shifts and line broadening, at two flare ribbons separating from each other, suggesting that chromospheric evaporation takes place in successively formed flaring loops throughout the flare. More importantly, we detect dominant blueshifts in the high temperature Fe {\sc xxi} line ($\sim$10 MK), in agreement with theoretical predictions. We also find that, in this flare, gentle evaporation occurs at some locations in the rise phase of the flare, while explosive evaporation is detected at some other locations near the peak of the flare. There is a conversion from gentle to explosive evaporation as the flare evolves.
\end{abstract}


\section{Introduction}
\label{intro}

In the standard solar flare model, magnetic reconnection takes place in the corona and releases a large amount of energy \citep{prie02}. The released energy is transported downward to the lower atmosphere by non-thermal particles and/or thermal conduction. Most of the energy deposits in the dense chromosphere and heats up the local plasma. Due to a pressure imbalance, the heated chromospheric plasma is driven up into the corona. This process is called chromospheric evaporation \citep{neup68,hira74,acto82}. Based on momentum balance, there are also some chromospheric materials moving downward, called chromospheric condensation \citep{fish87,canf90}. The evaporated hot plasma fills the newly formed flare loops that are then visible as post-flare loops in soft X-ray and extreme-ultraviolet (EUV) wavelengths.

Chromospheric evaporation can be detected in imaging and spectroscopic observations. \cite{silv97} reported that the footpoints of a flaring loop in hard X-ray and soft X-ray images from {\em Yohkoh} moved toward the loop-top source. \cite{liuw06} also found that the {\em RHESSI} hard X-ray sources in an M-class flare rose above the footpoints and merged into a loop-top source. Recently, \cite{nitt12} observed an evaporation upflow motion of several hundreds of km s$^{-1}$ in soft X-ray images from {\em Hinode}. There have been, however, only a few such imaging observations of chromospheric evaporation. In fact, our knowledge of chromospheric evaporation is mainly from spectroscopic observations. Mass flows related to evaporation can be detected by Doppler shift measurements in chromospheric and coronal lines: upward motions generate blueshifts and downward motions give rise to redshifts in the lines. Earlier studies reported blueshifts of 200--400 km s$^{-1}$ in the hot coronal Ca {\sc xix} line (e.g., \citealt{anto82,anto85,anto83,canf87,zarr88,wuls94,ding96}) using the Bent and Bragg Crystal Spectrometer (BCS) on board the {\em Solar Maximum Mission} ({\em SMM}; \citealt{acto80}) and {\em Yohkoh}/BCS \citep{culh91}. Similar blueshifts of 60--300 km s$^{-1}$ have also been obtained in the hot Fe {\sc xix} line (e.g., \citealt{teri03,teri06,bros04,delz06}) using the Coronal Diagnostic Spectrometer (CDS; \citealt{harr95}) on board the {\em Solar and Heliospheric Observatory} ({\em SOHO}). In the mean time, redshifts corresponding to downward velocities of tens of km s$^{-1}$ have been measured in the chromospheric and transition region (TR) lines (e.g., \citealt{wuls94,ding95,czay99,teri03,teri06,bros03,kami05,delz06}). Recently, both blueshifts and redshifts have been detected in different emission lines using the EUV Imaging Spectrometer (EIS; \citealt{culh07}) on board {\em Hinode} (e.g., \citealt{mill09,chen10,ying11,dosc13}).

According to the observed Doppler shifts, chromospheric evaporation can be divided into two types: gentle evaporation and explosive evaporation \citep{fish85a,fish85b,fish85c,mill06a,mill06b}. When the hot coronal lines are blueshifted and the lines formed in the upper chromosphere and TR are redshifted, it is considered as explosive evaporation. When all the coronal, transition regional, and chromospheric lines show blueshifts, it is referred to as gentle evaporation. These two types of evaporation are also distinguished by the speed of evaporating plasma: gentle evaporation is characterized by tens of km s$^{-1}$ (subsonic) upflows, while explosive evaporation characterized by hundreds of km s$^{-1}$ (supersonic) upflows (\citealt{zarr88} and references therein). Using hydrodynamic simulations, \cite{fish85a,fish85b} reported that there may exist an energy flux threshold of about 10$^{10}$ erg cm$^{-2}$ s$^{-1}$ (with a 20 keV energy cutoff) between gentle and explosive evaporations. However, recently, \cite{rubi15} found that stochastically accelerated electrons with an energy flux $<$10$^{10}$ erg cm$^{-2}$ s$^{-1}$ could also cause explosive evaporation. In addition, {\cite{reep15} claimed that the threshold between explosive and gentle evaporation is not a constant in the beam flux, but depends on the electron energy and duration of heating.} Note that, besides being driven by non-thermal electrons, both types of evaporation could also be driven by thermal conduction \citep{zarr88,mill08,batt09,ganw91,ganw92,liuw09}. In addition, both types of evaporation and the transition between them are observed in certain flare events \citep{bros09}, and explosive evaporation could appear not only in large flares (e.g., \citealt{mill06a,vero10}) but also in microflares \citep{bros10,chen10}.

Although a large number of observational results are consistent with chromospheric evaporation models, there still exist some discrepancies between observations and models. Evaporation models predict that hot emission lines, such as Fe {\sc xxiii}, Fe {\sc xix}, and Ca {\sc xix}, should exhibit dominant blueshifts. In observations, however, these lines show a dominant stationary component with only a weak blueshifted component indicating upflows of hundreds of km s$^{-1}$ \citep{anto82,ding96,mill09}. A few scenarios have been proposed to explain the observations: (1) the stationary plasma is from the top of flare loops where evaporated material is stopped \citep{dosc05}; (2) the plasma is moving in the direction perpendicular to the line of sight \citep{fale09}; (3) the stationary component comes from the loop strands in which the evaporation process has ceased, while the weak blueshifted component originates from the newly formed strands with ongoing evaporation, and the observed spectrum includes contributions of both components in the same field of view along the line of sight \citep{bros13}; (4) the resolution of current instruments is not high enough to observe the evaporating plasma at an early time \citep{dosc05}. Another problem concerns where the reversal from upflows to downflows takes place in explosive evaporation. It is usually considered that this reversal occurs at the site of energy deposition (e.g., \citealt{mill09,ying11}). Evaporation models predict that the reversal temperature is below 1 MK \citep{fish85a}, which was supported by some observations \citep{kami05,delz06}. However, higher reversal temperatures, such as 2 MK, or even 5 MK, were also found in observations \citep{mill08,ying11}. These observations pose challenges to evaporation models.

The recently launched Interface Region Imaging Spectrograph (IRIS; \citealt{depo14}) provides high resolution imaging and spectroscopic data in the UV wavelength range. Due to the observing modes, it needs some fortune for the spectrograph to capture a solar flare. Fortunately, IRIS has captured more than 200 flares even including a few X-class flares as of June 2015. Furthermore, for some flares, coordinated observations with another spectrometer, EIS, which provides high resolution spectroscopic data in the EUV wavelength range, are available. Combining the UV and EUV spectroscopic data obtained by these two instruments, we study the chromospheric evaporation process from the chromosphere and TR to the corona during an X-class flare. In Section \ref{observation}, we describe the flare observations; then we show our results of spectral analysis in Section \ref{result}; finally we give a summary and discussions in Section \ref{discussion}.

\section{Observations and Data Reduction}
\label{observation}

\begin{figure*}[htb]
\begin{center}
\includegraphics[width=16.5cm]{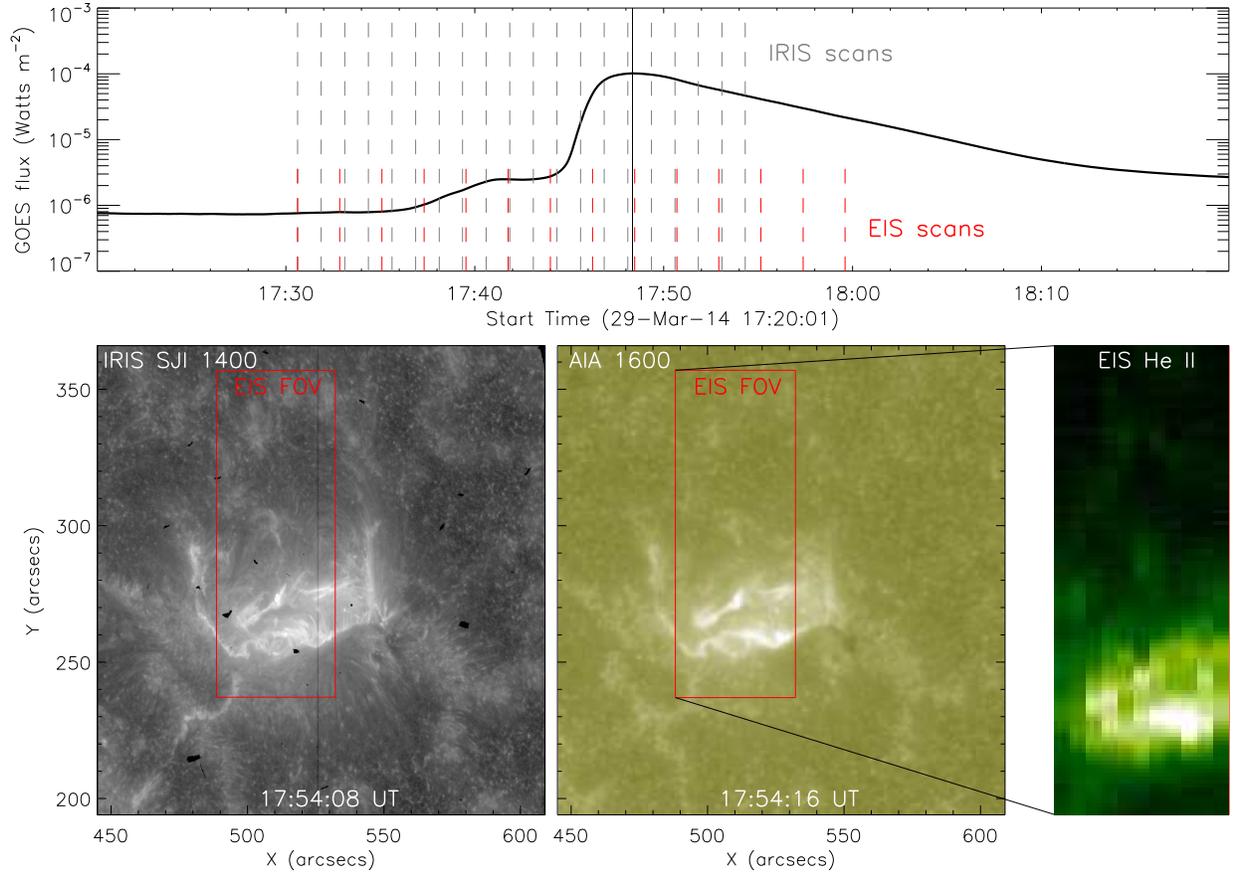}
\caption{Top panel: {\em GOES} 1--8 \AA~soft X-ray light curve for the X1.0 flare. The grey and red vertical dashed lines represent the multiple IRIS and EIS scans over the flaring region, respectively. Bottom panels: IRIS slit-jaw 1400 \AA~image (SJI 1400), AIA 1600 \AA~image, and EIS He {\sc ii} 256 \AA~intensity map several minutes after the {\em GOES} soft X-ray peak time. The red box shows the EIS FOV.}
\label{obs}
\end{center}
\end{figure*}

The flare analyzed here is an X1.0 flare that occurred on 2014 March 29 in NOAA AR 12017. This is the first X-class flare that IRIS has observed. Coordinated observations have also been obtained with EIS for this flare. Figure \ref{obs} shows a summary of the observations. The top panel presents the {\em GOES} 1--8 \AA~soft X-ray light curve, which peaks at 17:48 UT as indicated by the vertical solid line. The vertical dashed lines mark the times of multiple scans of IRIS and EIS over the flaring region. The bottom panels show the IRIS slit-jaw image in 1400 \AA, UV 1600 \AA~image by the Atmospheric Imaging Assembly (AIA), and EIS He {\sc ii} 256 \AA~intensity map. These images are taken at $\sim$17:54 UT, several minutes after the flare peak time. We can clearly see the two flare ribbons in the images. Note that the field of view (FOV) of IRIS overlaps with that of EIS.

IRIS obtains slit-jaw images and spectra in the near-UV (NUV, 2783--2834 \AA) and far-UV (FUV, 1332--1358 \AA~and 1389--1407 \AA) passbands using a slit with a width of 0.33\arcsec. The pixel scale is 0.167\arcsec. IRIS observed an area of 167\arcsec$\times$174\arcsec~in the slit-jaw 1400 \AA~images with a cadence of 26 s for the X1.0 flare. The slit scanned over an area of 14\arcsec$\times$174\arcsec~with 8 steps (coarse steps, implying a spatial resolution of 2\arcsec~across the slit) in each run that took 75 s. IRIS observed the active region for several hours till 17:54 UT. Here we use the data in the last 25 minutes (from 17:30 to 17:54 UT), which cover the rise phase through part of the decay phase of the flare. During this time period, the IRIS slit scanned over the flaring region 19 times, which are indicated by the grey dashed lines in Figure \ref{obs}. The IRIS level 2 data are used in our study. The data have been processed with dark current subtraction, flat field correction, and geometric and wavelength corrections. For the spectra, the position of spectral lines suffers an orbital variation due to the temperature change of the detector and the change of spacecraft--Sun distance over the orbit. This line center variation was eliminated through the routine {\em iris\_orbitvar\_corr\_l2.pro} in the standard IRIS software data package. The IRIS lines studied here are listed in Table \ref{line}.

EIS acquires EUV spectra in the short-wavelength range of 170--210 \AA~and long-wavelength range of 250--290 \AA~with a spectral resolution of 0.0223 \AA~pixel$^{-1}$. The 1\arcsec~slit of EIS was used to scan over the flaring region with a step of 4\arcsec. There are 11 steps in each run that took 2 minutes and 14 s, with a scanning area of 44\arcsec$\times$120\arcsec~that overlaps with the IRIS FOV. EIS scanning repeated 13 times from 17:30 UT until the end of the coordinated observations with IRIS. The EIS data are reduced using the standard EIS software data package, with corrections for detector bias, dark current, hot pixels, and cosmic ray hits. The tilt of the slit is also corrected. In addition, we make a correction for the line position shift due to temperature variation of the spectrometer. The EIS lines used here are also listed in Table \ref{line}.

To measure Doppler velocity, we use average line center to determine the reference wavelength of spectral lines. For the high temperature lines, e.g., Fe {\sc xxi} and Fe {\sc xxiii}, which cannot be seen before the flare onset, we use the average line center over a loop-top region in the decay phase of the flare as the reference wavelength by assuming that the plasma in this region has reached a static state without significant mass flows well after the flare peak. With this method, we find the reference wavelength for the Fe {\sc xxi} line to be 1354.126 \AA, which is consistent with the value 1354.106$\pm$0.023 \AA~in \cite{youn15}. Alternatively, we can also use the relatively cool lines in the same spectral window, namely O {\sc i}, to calculate the reference wavelength of Fe {\sc xxi}. This assumes that the cool O {\sc i} line is at rest before the onset of the flare; then we find the offset between the average line center of O {\sc i}, observed in quiescent regions by IRIS, and the theoretical line center given in the CHIANTI database, and apply this same offset to the Fe {\sc xxi} line. For comparison, this method yields the rest wavelength of Fe {\sc xxi} to be 1354.124 \AA, which is very close to the value obtained in the former method. For the other warm and cool lines, i.e., Fe {\sc xvi}, Fe {\sc xii}, Si {\sc iv}, He {\sc ii}, C {\sc ii}, and Mg {\sc ii}, we use the average line center before the flare onset as the reference wavelength. The uncertainty in the measured velocities for all the lines is estimated to be less than 10 km s$^{-1}$ considering the spectral resolution and calibration errors of IRIS  and EIS data.

Co-alignment between IRIS and EIS images is made as follows. We first co-align the IRIS slit-jaw 1400 \AA~image with the AIA 1600 \AA~image, both of which show similar features (see Figure \ref{obs}). Then we co-align the AIA 304 \AA~image, in which the emission is mainly from the He {\sc ii} line, with the EIS He {\sc ii} intensity map. In addition, we also correct offsets (mainly in the Y-direction) between different spectral windows of EIS itself. Considering the spatial resolution of IRIS and EIS, namely 2\arcsec$\times$0.167\arcsec~and 4\arcsec$\times$1\arcsec, respectively, the accuracy of our co-alignment between IRIS and EIS is estimated to be about 4\arcsec$\times$4\arcsec.

\section{Analysis and Results}
\label{result}

\subsection{Analysis of the IRIS UV spectra}

IRIS provides high resolution spectra in the FUV and NUV passbands, most of which were not well studied for solar flares before. In the following, we first present the spectra along the slit (Figure \ref{iris_spec}) and the line profiles at flare ribbon pixels (Figures \ref{profile_n0}--\ref{profile_s3}) observed by IRIS to show its capability of diagnosing chromospheric evaporation. Then we present the evolution of Doppler velocity along the slit (Figure \ref{iris_map}) to show the temporal variation of chromospheric evaporation at the two flare ribbons. Here we analyze the hot Fe {\sc xxi} line formed at $\sim$10 MK, the Si {\sc iv} line formed in the TR, and the Mg {\sc ii} and C {\sc ii} lines formed in the upper chromosphere. Note that the C {\sc ii} line could also be formed in the low TR. 

Figure \ref{iris_spec} shows the slit-jaw 1400 \AA~images and the Fe {\sc xxi}, Si {\sc iv}, C {\sc ii}, and Mg {\sc ii} h\&k spectra along the IRIS slit at different scan steps and different evolution stages of the flare. The top panels give the slit-jaw image and the spectra at the seventh step with the position indicated by the white vertical line in the slit-jaw image just two minutes before the flare peak time. From the image, one can clearly see the two flare ribbons which are cut across by the IRIS slit. Around both ribbons, the hot Fe {\sc xxi} line emission appears and shows some significant blueshifts, while the other cool lines show a strong red-wing enhancement. Note that in the brightest part of the ribbons, the Si {\sc iv} spectra and the slit-jaw image are saturated. A careful inspection also reveals existence of a large number of narrow lines in all the spectral windows, especially near the flare ribbon. In the middle panels, we plot the slit-jaw image and the spectra at the sixth step of the slit at the flare peak time. One can see that there appears a slight separation between two flare ribbons compared with the earlier time. Near the south ribbon, the Fe {\sc xxi} line shows very strong blueshifts of more than 200 km s$^{-1}$, and the velocity decreases quickly to nearly zero toward the north that is close to the flare loop-top region. However, the Fe  {\sc xxi} line intensity of the loop-top is much greater than that near the ribbon. By comparison, at flare ribbons, the cool Si {\sc iv}, C {\sc ii}, and Mg {\sc ii} lines show a red-wing enhancement. The bottom panels show the slit-jaw image and the spectra at the third step in the decay phase of the flare. One can see that the two flare ribbons have a further separation in opposite directions. One can also find some short post-flare loops in the south around the slit as indicated by the white vertical line in the slit-jaw image. It is seen that the spectra in the loop-top region show some particular features: the hot Fe {\sc xxi} line is almost unshifted though its intensity is greatly enhanced, while the cool lines exhibit some broadening, which may be due to superimposed line-of-sight velocities of cooling plasma.

The general result revealed in Figure \ref{iris_spec} is that the hot Fe {\sc xxi} line shows significant blueshifts and the other cool lines show a red-wing enhancement at the flare ribbons. This can be considered as clear evidence of chromospheric evaporation, in particular, explosive evaporation. To display the evaporation signatures during the flare more clearly, we plot the line profiles at four ribbon pixels (named as N0, S1, S2, and S3) marked by the red bars in the slit-jaw images (Figure \ref{iris_spec}), and measure the Doppler velocities. When measuring the Doppler velocity, we adopt different methods for different spectral lines. For the optically thin Fe {\sc xxi} and Si {\sc iv} lines, we use Gaussian function to fit the observed line profiles. Since there are several narrow and cool lines (shown in the top left panel of Figure \ref{profile_n0}, `u' means unidentified) blending with the Fe {\sc xxi} line in general, we adopt a multi-Gaussian function with nine components representing nine lines and assume that the lines from the same ion having the same Doppler velocity and width in order to fit the Fe {\sc xxi} profiles. However, a double Gaussian function is applied to the line profiles that exhibit only C {\sc i} blending with Fe {\sc xxi}, such as the ones near the flare loop top. Most of the Si {\sc iv} line profiles show a good single Gaussian shape, and we adopt a single Gaussian function to fit the Si {\sc iv} line. However, there are some Si {\sc iv} line profiles showing an asymmetry, mainly at the flare ribbon, so we also use a double Gaussian function to fit those line profiles and compare with the fitting results by a single Gaussian function. It shows that the velocity derived from a double Gaussian fitting (the redshifted component) is a little larger than the velocity from a single Gaussian fitting, but it does not change the sign of the velocity. The C {\sc ii} and Mg {\sc ii} lines are optically thick, which are formed in the chromosphere through a complex radiative transfer process. The core (h$_3$ or k$_3$ component) of the Mg {\sc ii} h\&k resonance lines is formed in the upper chromosphere just below the TR, whereas the wings (h$_2$ and k$_2$) are formed lower. In the quiet Sun (and pre-flare), the Mg {\sc ii} resonance lines usually exhibit a central reversal at the core. In this flare, the central reversal is not present in most of the observed Mg {\sc ii} h\&k resonance lines as well as the C {\sc ii} lines at the flaring locations. The line profiles exhibit a single emission peak, which appears to be redshifted with respect to the quiescent line center. Here we adopt a bisector or moment analysis method to estimate the line center shift of C {\sc ii} and Mg {\sc ii}. Because of a similarity between the two Mg {\sc ii} (or C {\sc ii}) resonance lines, we only present the results for one of the lines. The observed line profiles and the fitting results at the selected four ribbon pixels are shown in Figures \ref{profile_n0}--\ref{profile_s3}.

In Figure \ref{profile_n0}, we plot the line profiles of Fe {\sc xxi}, Si {\sc iv}, C {\sc ii}, and Mg {\sc ii} k, together with the fitting curves or net emissions, at the north ribbon pixel N0 that brightened in the rise phase of the flare. It is seen that the fitting curve of the Fe {\sc xxi} line matches the observed one very well. The Fe {\sc xxi} line exhibits a significant blueshift with a Doppler velocity of 117 km s$^{-1}$. The Si {\sc iv} line, however, shows a red asymmetry. When using a single Gaussian function to fit the line profile, we get a redshift with a velocity of 15 km s$^{-1}$. If using a double Gaussian fitting, we can obtain a larger redshift velocity of 55 km s$^{-1}$ with an intensity ratio of 0.6. Similarly, the C {\sc ii} line shows a redshift with a velocity of $\sim$36 km s$^{-1}$ by a bisector method (at $\sim$40\% of the peak intensity) or 37 km s$^{-1}$ by the moment analysis on the flare net emission (subtracting the pre-flare emission). The Mg {\sc ii} k line also displays a red asymmetry, seen from the moment analysis on its pre-flare ($+$5 km s$^{-1}$), during-flare ($+$14 km s$^{-1}$), and flare-net ($+$30 km s$^{-1}$) emissions, respectively.

Figure \ref{profile_s1} shows the line profiles and fitting curves or net emissions at the south ribbon pixel S1 that brightened also in the rise phase of the flare. The intensity at this pixel is much stronger than that at pixel N0. Specifically, the intensity of Fe {\sc xxi} at S1 is three times the one at N0, and the Si {\sc iv} and C {\sc ii} lines seem saturated at S1 (here we analyze the unsaturated Si {\sc iv} and C {\sc ii} line profiles at the closest ribbon pixel nearby S1 instead). In addition, the Mg {\sc ii} lines show a stronger emission at S1 than at N0. At S1, we obtain a blueshift with a velocity of 64 km s$^{-1}$ in the hot Fe {\sc xxi} line, and some redshifts or a red asymmetry with velocities of tens of km s$^{-1}$ in the cool Si {\sc iv}, C {\sc ii}, and Mg {\sc ii} lines.

Figure \ref{profile_s2} presents the line profiles and fitting curves or net emissions at the south ribbon pixel S2 that brightened at the flare peak time. At this pixel, the Si {\sc iv} and C {\sc ii} lines are also saturated, and similarly, we analyze the unsaturated profiles nearby instead. The intensity of Fe {\sc xxi} at S2 is not as strong as the one at S1, but the Doppler velocity at S2 reaches as high as 214 km s$^{-1}$, much larger than that at S1. In addition, the cool lines also exhibit redshifts with velocities of 20--45 km s$^{-1}$. The Mg {\sc ii} line also show a broad line profile. All of these indicate that the evaporation is very strong at this pixel in the impulsive phase.

In Figure \ref{profile_s3}, we display the line profiles and fitting curves or net emissions at the south ribbon pixel S3 that brightened in the decay phase of the flare. It is seen that the intensities of all the lines at S3 have significantly decreased compared with the line intensities at the previous pixels in the rise and impulsive phases. The Fe {\sc xxi} line also shows a decreased blueshift velocity of 30 km s$^{-1}$. The other cool lines still show some red asymmetry.

The high temperature Fe {\sc xxi} line, however, is symmetric and well fitted by a single Gaussian function with a blueshift velocity. This suggests that, on the flare ribbon, the Fe {\sc xxi} line is blueshifted as a whole. In particular, at S2 in the impulsive phase, the line shows a blueshift with a strong velocity of 214 km s$^{-1}$. This is different from previous results, which only show a weak blueshifted component superimposed on a dominant stationary component (e.g., \citealt{ding96,mill09}). Our results are consistent with the prediction by theoretical models that hot plasma is massively driven upward in an impulsively heated flare loop. It is likely that the high spatial resolution of IRIS allows us to resolve the locations where energy deposition and strong dynamics take place in the atmosphere, whereas previous observations with low spatial resolutions may reflect a mixture of unresolved multiple loops with both blueshifted plasma and stationary plasma.

Besides the line profiles at ribbon pixels, we also analyze the spectra at all the pixels along the slit. Figure \ref{iris_map} shows the time sequence of Doppler velocity, as well as non-thermal velocity derived from the line width, of the Fe {\sc xxi} and Si {\sc iv} lines, for pixels along the slit between the two green bars indicated in the bottom left panel of Figure \ref{iris_spec}. The map covers a spatial scale of 30\arcsec~along the slit, which lies across the two ribbons separating from each other. Since the Fe {\sc xxi} line emission is weak in many pixels especially at the north ribbon, to get a better signal-to-noise ratio, we rebin the data to sum up spectra in six IRIS pixels along the slit. We have confirmed that rebinning the data does not affect significantly the measured Doppler shift distribution along the slit. Now the pixel size is 0.33$\arcsec$ by 1$\arcsec$ in Figure \ref{iris_map}. Also note that, the black data points in the Fe {\sc xxi} velocity panel represent the strong Doppler velocities that are far beyond the colorbar range ($\pm$150 km s$^{-1}$). Some of these velocities even exceed 250 km s$^{-1}$. In addition, in Figure \ref{iris_map} we only show the time sequence of velocity along the slit for the third step. For the other steps, either the observed data contain bad points (with all the intensity counts being a negative value of -200), or the Fe {\sc xxi} emission is too weak to be identified at the north ribbon.

More interestingly, we find an evident pattern of ribbon separation in the velocity maps, as shown in Figure \ref{iris_map}. The separation of two flare ribbons is an indication of progressive reconnection that forms flare loops successively and deposits energy at the feet of these newly formed loops. In this event, the Fe {\sc xxi} velocity map shows a pair of blueshift fronts separating from each other at an apparent speed of about 20 km s$^{-1}$, the same as the ribbon separation speed. These blueshifts, corresponding to upflow velocities in the range of 50--300 km s$^{-1}$, are located at the bright fronts of flare ribbons, as indicated by the intensity contours of the Mg {\sc ii} h (or Si {\sc iv}) line. The Fe {\sc xxi} velocity map also reveals that, on the ribbons, particularly the south ribbon, the region with significant blueshifts extends over a spatial scale of 2--3 arcseconds along the slit. At the same location, the blueshift diminishes in several minutes. In the south ribbon, after the bright front passes, the blueshift turns into a redshift (with a velocity of 3--15 km s$^{-1}$). Most of the redshift velocities are within the measurement uncertainty. It is likely that some of the redshift signatures are real, though we are not certain what might be the reason for the redshift in the hot Fe {\sc xxi} line. 

Similar evolution patterns of the dynamics on the flare ribbon fronts are also observed in other velocity maps. In particular, significant redshifts, with a velocity of a few tens of km s$^{-1}$, in the cool Si {\sc iv} line are observed at locations of strong ribbon emission. (Note that the persistent weak blueshift below the north ribbon may not be related to the flare.) At these places, the maximum line broadening in both the Fe {\sc xxi} and Si {\sc iv} lines is also observed to be of the order of 100 km s$^{-1}$ in terms of non-thermal velocity. These large non-thermal velocities on the flare ribbons may imply multiple velocity components in unresolved strands when energy deposition takes place. In fact, \cite{mill11} has found a good correlation between non-thermal velocity and Doppler velocity in the Fe {\sc xv} and Fe {\sc xvi} lines at loop footpoints, and the author interpreted the excess broadening as a superposition of flows (turbulence), presumably caused by chromospheric evaporation. The dynamic features diminish between the two ribbons.

These results provide clear evidence of chromospheric evaporation at the sites where impulsive energy deposition takes place. The time scale of the evaporation signature (particularly the blueshift) at one site is estimated to be 1--2 minutes (might be constrained by the observing cadence), and along the slit, the spatial scale of significant blueshift signatures is 2--3 arcseconds. The presence of coincident blueshift in the hot line and redshift in the cool lines suggests the regime of explosive evaporation. This evaporation process proceeds well into the decay phase of the flare, indicative of continuous energy release/deposition from the start to the decay phase \citep{mcti93,jian06}. Evaporation occurs in different loops formed successively by reconnection, therefore, it exhibits an apparent motion outward with a similar velocity to the separation velocity of flare ribbons.

\subsection{Analysis of the EIS EUV spectra}

\begin{figure*}[htb]
\begin{center}
\includegraphics[width=16cm]{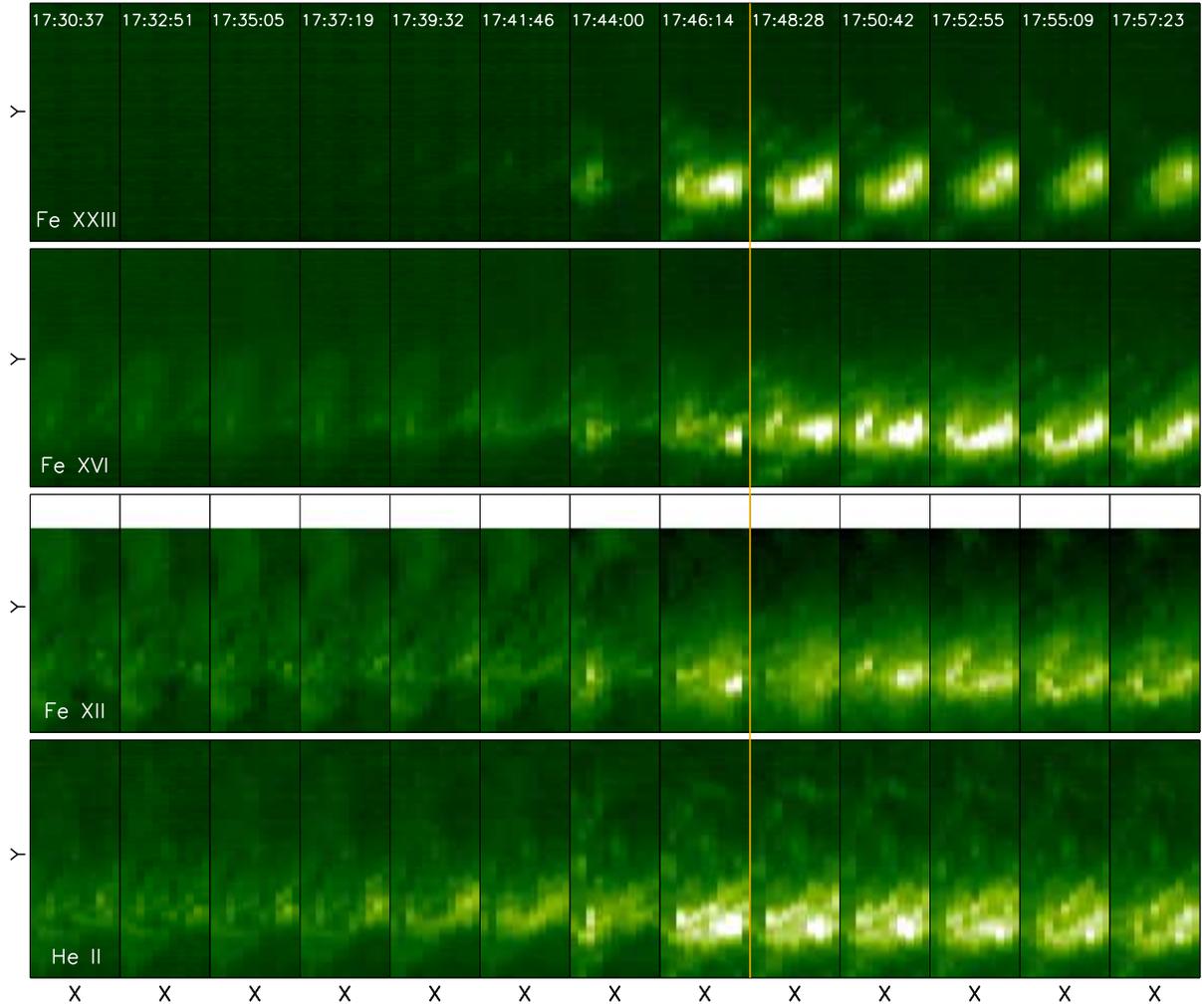}
\caption{Evolution of the intensity in four EIS spectral lines. The yellow line marks the {\em GOES} soft X-ray peak time. The time on the top row shows the starting time of each scan. All the intensity panels have the same FOV, as shown by the red box in Figure \ref{obs}.}
\label{eis_int}
\end{center}
\end{figure*}

Aside from the Fe {\sc xxi} line ($\sim$10 MK), IRIS observations are mostly made in chromospheric and transition region lines with characteristic temperatures up to 0.1 MK. In comparison, EIS provides EUV spectra that focus on diagnosing properties of transition region and corona plasma of temperatures from 0.05 MK to dozens of MK. To have a comprehensive picture of atmospheric dynamics from the chromosphere to the corona, we also select four emission lines from EIS, listed in Table \ref{line}, to study the chromospheric evaporation process in the X1.0 flare. These lines, He {\sc ii}, Fe {\sc xii}, Fe {\sc xvi}, and Fe {\sc xxiii}, form at 0.05, 1.3, 2.5, and 13 MK, respectively. Figure \ref{eis_int} shows the intensity map in these four lines. No emission is observed in the very hot Fe {\sc xxiii} line during the early phase of the flare. At 17:46 UT, the start of the major GOES soft X-ray peak, this line shows significant emission in the flaring region. The intensity maps of the other cooler lines also exhibit two flare ribbons, with the south ribbon much stronger than the north ribbon. In the intensity map of the coolest He {\sc ii} line, in particular, one can notice separation of the two ribbons, which is consistent with IRIS observations. 

\begin{figure*}[htb]
\begin{center}
\includegraphics[width=16cm]{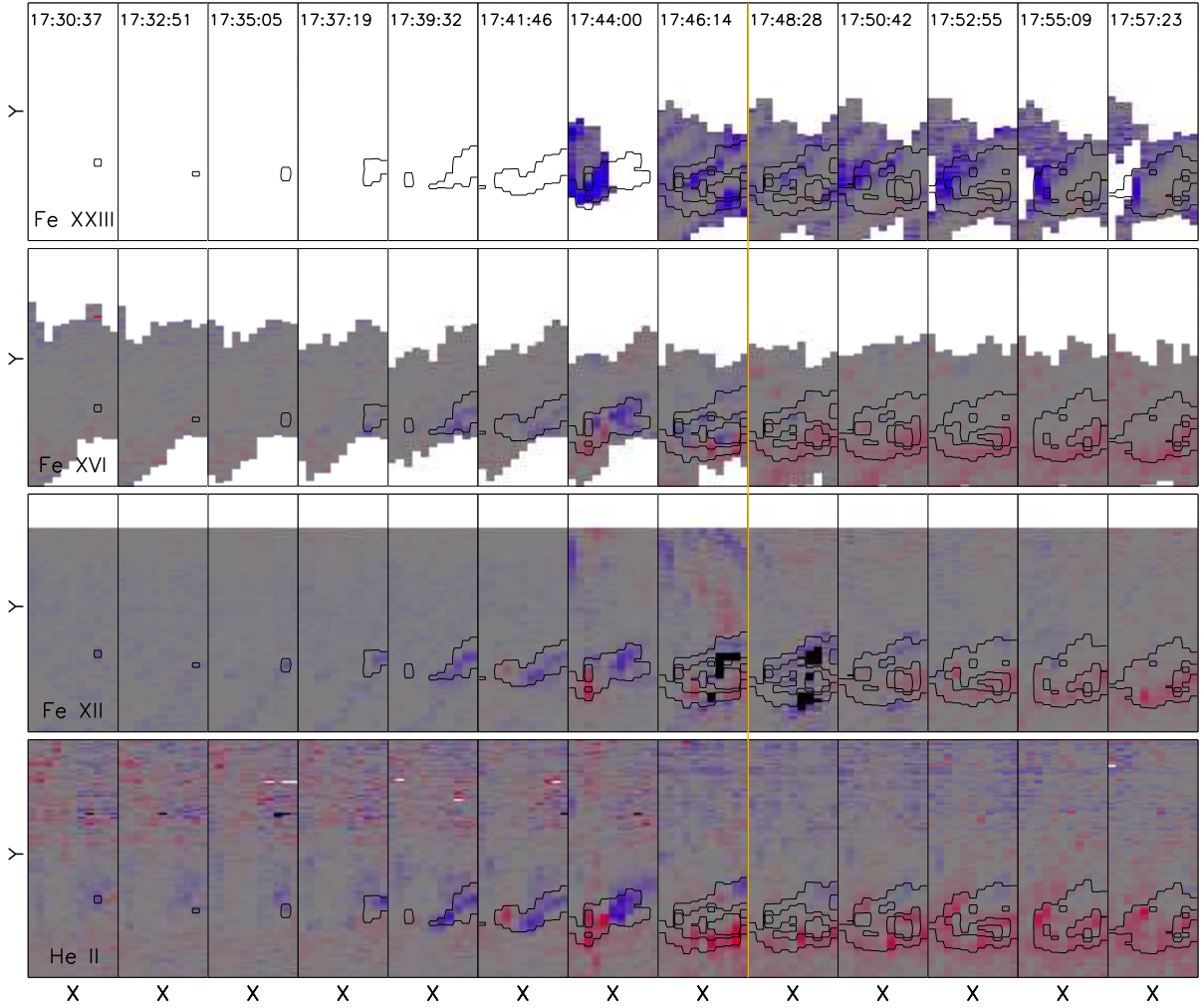}
\caption{Evolution of the Doppler velocity in four EIS spectral lines, superimposed on the intensity contours of the He {\sc ii} line. The velocity range is $\pm$ 200 km s$^{-1}$. The yellow line marks the {\em GOES} soft X-ray peak time. The time on the top row shows the starting time of each scan. The blank part in the Fe {\sc xxiii} and Fe {\sc xvi} maps means no values, since the signal-to-noise ratio at those pixels is too low to get reliable fitting results. The dark pixels in the Fe {\sc xii} velocity map represent a failed fitting to derive the Doppler velocity due to the Fe {\sc xii} line submerged in the wing of the Fe {\sc xxiv} line that becomes strong and broad around the flare peak.}
\label{eis_vel0}
\end{center}
\end{figure*}

In Figure \ref{eis_vel0}, we present the Doppler velocity map of the four EIS lines constructed from multiple scans. Here we only show the results for the pixels that have enough signal-to-noise ratio (mainly for the Fe {\sc xxiii} and Fe {\sc xvi} lines). All the velocities are derived by fitting the observed line profiles to a single Gaussian function. From the Doppler velocity map, we can see that the hot Fe {\sc xxiii} line shows strong blueshifts at the flare ribbons. The relatively cooler lines of Fe {\sc xvi}, Fe {\sc xii}, and He {\sc ii} show a variety of blueshifts before the flare peak time. Note that some strong buleshifts in these cooler lines may be related to a filament eruption (for details see \citealt{klei15}) during the early time of the flare. From 17:46 UT, two minutes before the flare peak, on the flare ribbons, the cooler lines become mostly redshifted, while the hot Fe {\sc xxiii} line stays blueshifted. These manifest explosive evaporation, which confirms the IRIS results. Note that, in the velocity map of the hot Fe {\sc xxiii} line, there also appear some patches with redshifts near the south ribbon, which might be caused by the uncertainty in velocity measurements.

\begin{figure*}[htb]
\begin{center}
\includegraphics[width=16cm]{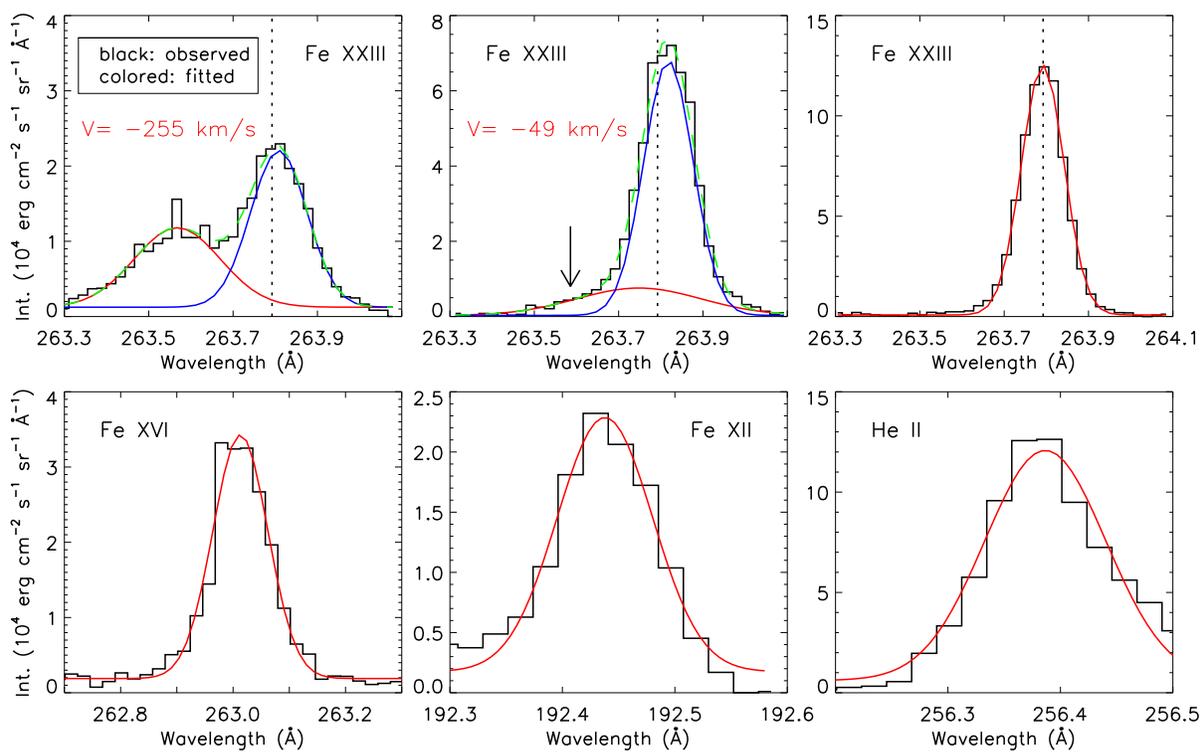}
\caption{Sample line profiles (black curves) of four EIS emission lines observed during the flare, with the Gaussian fitting shown in colored curves. Note that, for the Fe {\sc xxiii} line in the top panels, we use either a single (right) or a double (left and middle) Gaussian function to fit the observed line profiles. The arrow marks a very weak blue asymmetry in the Fe {\sc xxiii} line. The vertical dotted lines indicate the reference wavelength of Fe {\sc xxiii}.}
\label{eis_prof}
\end{center}
\end{figure*}

We note that almost all of the Fe {\sc xvi}, Fe {\sc xii}, and He {\sc ii} line profiles show a single peak, and can be well fitted with a single Gaussian function (see bottom panels in Figure \ref{eis_prof}). However, some of the Fe {\sc xxiii} line profiles exhibit a weak blue asymmetry or even double peaks (mainly at the flare ribbon), though the majority still show a single peak of Gaussian shape (especially near the loop-top region). These different types of Fe {\sc xxiii} line profiles observed during the flare are presented in the top panels of Figure \ref{eis_prof}. In the following, we alternatively adopt a double Gaussian function to fit the Fe {\sc xxiii} line profiles that display double peaks or weak blue asymmetry, as shown in the top left and middle panels of Figure \ref{eis_prof}.

Figure \ref{eis_tmap} shows the evolution of Doppler velocities along the slit in the four EIS lines. The EIS slit position (its eighth step) is nearly the same as the IRIS slit position (its third step) shown in Figure \ref{iris_map}. Due to the different spatial resolutions as well as different observing times, there may be a relative shift of several arcseconds in both X and Y directions between the EIS slit and IRIS slit. In addition, the time scales of EIS and IRIS scans are different: EIS takes a longer time than IRIS for each scan, and EIS observations persist five minutes longer before the end of the flare. These maps also show an apparent separation motion between two ribbons exhibiting Doppler shifts. However, the pattern is not as evident as the IRIS Fe {\sc xxi} measurement in Figure \ref{iris_map}, which is likely due to the low spatial and temporal resolutions of EIS. Note that, the Fe {\sc xxiii} Doppler velocity map is derived from a double Gaussian fitting to the line profiles displaying double peaks (within the red rectangular box in the panel) or a week blue asymmetry, and the velocities represent the shifts of blueshifted components, which can exceed 200 km s$^{-1}$. The mix of the blueshifted and stationary components in the EIS Fe {\sc xxiii} line, most likely due to the lower spatial resolution of EIS, explains why the blueshift signature is not as prominent as demonstrated in the IRIS Fe {\sc xxi} line.

\subsection{Doppler Velocity versus Temperature in Multiple UV and EUV Lines}

Combining the multi-temperature UV and EUV lines from IRIS and EIS, we can study Doppler shift signatures in cool lines and hot lines to distinguish different types of chromospheric evaporation at different locations during the evolution of the X1.0 flare. We select four time-space locations marked in Figure \ref{eis_tmap}. For each location, the spectra is from an area of 4\arcsec$\times$4\arcsec. Three locations (1, 2, and 3) are on the flare ribbons and brightened before the peak of the flare. These places show prominent evaporation signatures. At the fourth location, evaporation has stopped after the peak of the flare. Doppler velocities in different lines are plotted as a function of temperature in Figure \ref{eis_iris}. It is seen that, at location 1, all the spectral lines with different temperatures are blueshifted, which can be considered as gentle evaporation. At locations 2 and 3, the relatively hot lines are blueshifted, while the cool lines are redshifted, which is regarded as explosive evaporation. Note that, at locations 2 and 3, the velocity in the Fe {\sc xxiii} line marked by a vertical bar is derived from the blueshifted component, when a double Gaussian fitting is applied. The plots also suggest that, in the event of explosive evaporation, the reversal temperature from upflows to downflows is likely below 1 MK at location 2 and close to 1 MK at location 3, although there is a large uncertainty due to lack of spectral lines covering the temperature range from 0.1 to 1 MK. These results are similar to previous studies (\citealt{kami05,mill08}). Different flow reversal temperatures would imply that the energy deposition height, rate, or duration may be different at these two locations \citep{liuw09}. 

These four locations are at different positions along the slit (location 4 overlaps partly with location 1). Their Doppler velocities plotted in Figure \ref{eis_iris} are for different times at different stages of the flare. For locations 1, 2, and 3, chromospheric evaporation is in progress. Evaporation at location 1 is gentle evaporation which begins at the start of the major peak of the flare. Evaporation at locations 2 and 3 is explosive evaporation which occurs about two minutes later, close to the peak time of the flare. Therefore, in the flare, different locations can show different types of evaporation, and the evaporation can vary from a gentle one to an explosive one as the flare evolves. These results indicate that energy deposition, reflected by the magnitude and types of chromospheric evaporation, varies with time and space during a flare. In more details, there are two possible reasons for the time- and space-dependent energy deposition: (1) the injected energy flux of the electron beam or conduction front that deposits the energy is time- and space-dependent; and (2) the magnetic and density structure of the atmosphere, which determines heating and dynamics of the lower atmosphere, may change from place to place.

\section{Summary and Discussions}
\label{discussion}

We have studied the chromospheric evaporation process during an X1.0 flare combining the IRIS UV and EIS EUV spectroscopic data. From the IRIS data with high temporal and spatial resolution and a good time coverage of the flare, we have detected dominant blueshifts in the high temperature Fe {\sc xxi} line ($\sim$10 MK) at the flare ribbons. The Fe {\sc xxi} line profile at most locations on flare ribbons is symmetric and very well fitted by a single Gaussian function, and the entire line is blueshifted. The dominant blueshifts throughout the flare have rarely been observed by other spectrographs. Therefore, our result provides an observational support to theoretical models predicting that high temperature lines should be entirely or mostly blueshifted. 

We have constructed Doppler velocity maps derived from multiple lines obtained by IRIS and EIS slits at multiple times during the rise, peak, and decay of the flare. We have found that blueshifts in the high temperature Fe {\sc xxi} line and redshifts in other cooler lines from IRIS are located at the bright fronts of two flare ribbons, and spread out in the same manner as the separation of two ribbons. This evaporation process is seen to proceed into the decay phase of the flare. In addition, line broadening is also observed at flare ribbons. These results confirm that dramatic dynamic processes occur in the flaring atmosphere when energy deposition takes place. The pattern of ribbon separation, together with the dynamic features, reflects continuous magnetic reconnection and energy release. Similar evolution patterns are observed in velocity maps constructed with the EIS EUV spectra, though with a much lower spatial (and temporal) resolution.

By analyzing multiple lines observed by IRIS and EIS, we also examine Doppler shifts in cool lines and hot lines to study different types of chromospheric evaporation. In this flare, the evaporation process shows temporal and spatial variations: gentle evaporation is observed at one location during the rise of the flare, and explosive evaporation is observed at some other locations several minutes later, close to the peak of the flare. The conversion from gentle to explosive evaporation during the X1.0 flare may be related to non-thermal electron beams. We have examined the X-ray light curves from RHESSI, which show evident non-thermal hard X-ray emission up to 100 keV. The non-thermal hard X-ray light curve peaks at 17:46 UT, close to the scanning time at locations 2 and 3, both of which show explosive evaporation. By comparison, location 1 shows gentle evaporation that occurs two minutes earlier when high energy hard X-ray emission has not been enhanced.

Using high resolution spectra from IRIS, we find dominant blueshifts in the high temperature Fe {\sc xxi} line. The evaporation model predicts that the high temperature lines should be entirely or mostly blueshifted. However, there have been very few spectral observations of such kind to support the model prediction. In the past, Doppler shift measurements were obtained in high temperature lines such as Ca {\sc xix} by BCS and Fe {\sc xix} by CDS, respectively, with low spatial resolution. These measurements showed a weak blueshifted component superimposed on a dominant stationary component. These earlier observations could not reveal a dominant blueshift component, perhaps for two reasons. First, dominant blueshifts in high temperature lines usually appear in the early evaporation phase \citep{dosc05} when the high temperature emissions are very weak. Secondly, these dynamic features may occur in a very small region that could not be resolved with past instrument capabilities, and the observed spectra included contributions from nearby regions. Recently, the Fe {\sc xxiii} and Fe {\sc xxiv} lines with high spatial resolution from EIS were used to diagnose mass motions, and it was found that these high temperature lines show a dominant blueshifted component \citep{wata10,youn13} or that these lines are blueshifted as a whole \citep{bros13}. Nevertheless, such observations were too few to be conclusive. For example, \cite{bros13} found the above  phenomenon appears at one location and lasts for only about two minutes. Here, we use the Fe {\sc xxi} line from IRIS to study the evaporation process. We find that the line is blueshifted as a whole (also reported in different flares by \citealt{tian14b,tian15,youn15,poli15,grah15} recently) in the early evaporation phase in multiple loops continuously formed during the flare, thanks to the high temporal and spatial resolution of IRIS, as well as its high sensitivity in detecting high temperature emissions in the early phase of flare heating. 

From the high resolution spectra by IRIS, we have observed a dynamically evolving evaporation pattern characterized by blueshifts in high temperature ($\sim$10 MK) lines, which are co-spatial and coincident with separating flare ribbons. Flare ribbon separation has been observed for many years in H$\alpha$, UV, EUV, as well as hard X-ray images. However, previous studies have rarely reported separation of flare ribbons accompanied by coincident significant blueshifts, in particular, in high temperature lines. Flare atmospheric dynamics have been extensively studied; most previous studies have reported time evolution of chromospheric evaporation at some locations or spatial distribution of Doppler velocities at some given times (e.g., \citealt{czay99,bros03,bros04,bros09,bros10,bros13}). However, as limited by the time cadence and spatial resolution of the spectrometers, these past studies have not reported the spatio-temporal variation of the velocity pattern as shown in this work. Our result thus provides novel evidence of flare dynamics, which, together with what have been found in previous studies, can be explained in terms of the standard flare model: magnetic reconnection occurs successively at growing altitudes forming new and longer flare loops, and separating ribbons outline the feet of these successively formed loops, along which energy flux is deposited to heat the lower atmosphere and generate chromospheric evaporation at the feet; this process results in apparent separation of the velocity ribbons seen in the Dopplergrams. Note that reconnection proceeds continuously into the decay phase of the flare.  

We also note that the time and spatial scales of the strong blueshifts (in the hot Fe {\sc xxi} line) and redshifts (in the cool Si {\sc iv} line) at the ribbon are 1--2 minutes and 2--3 arcseconds, respectively. In general, these scales are a result of convolution of several factors: (1) the spread speed of flare ribbon (related to reconnection rate): the faster the spread motion is, the wider the blueshift band would be; (2) the duration of energy deposition at the loop footpoint, immediately after energy release by reconnection; (3) the hydrodynamic time scale within the flare loop; and (4) temporal and spatial resolutions of observations. Studying the ribbon dynamic features as revealed by emission features at different layers would help us to understand the flare heating source, rate, and duration, and further provide constraints to loop heating models and numerical simulations of flares. Note that the temporal and spatial scales of blueshifts/redshifts might be different in different flares.

Besides high temperature lines, IRIS provides important chromospheric and TR lines that can help us diagnose the lower atmospheric condition during chromospheric evaporation. These lines include the Mg {\sc ii} h\&k, C {\sc ii}, and Si {\sc iv} resonance lines. The Mg {\sc ii} and C {\sc ii} lines are optically thick, and formed in the chromosphere through a complex radiative transfer process. Interpretation of these optically-thick lines requires sophisticated methods such as through radiative transfer modeling. In the present study, we adopt a bisector or moment analysis method on the flare net emission to estimate the plasma motion in the chromosphere. Our results provide an estimate of the velocity field in the relatively low layers (i.e., the upper chromosphere), in comparison with the velocities in the transition region and corona deduced with optically-thin lines.

\begin{figure*}[htb]
\begin{center}
\includegraphics[width=15cm]{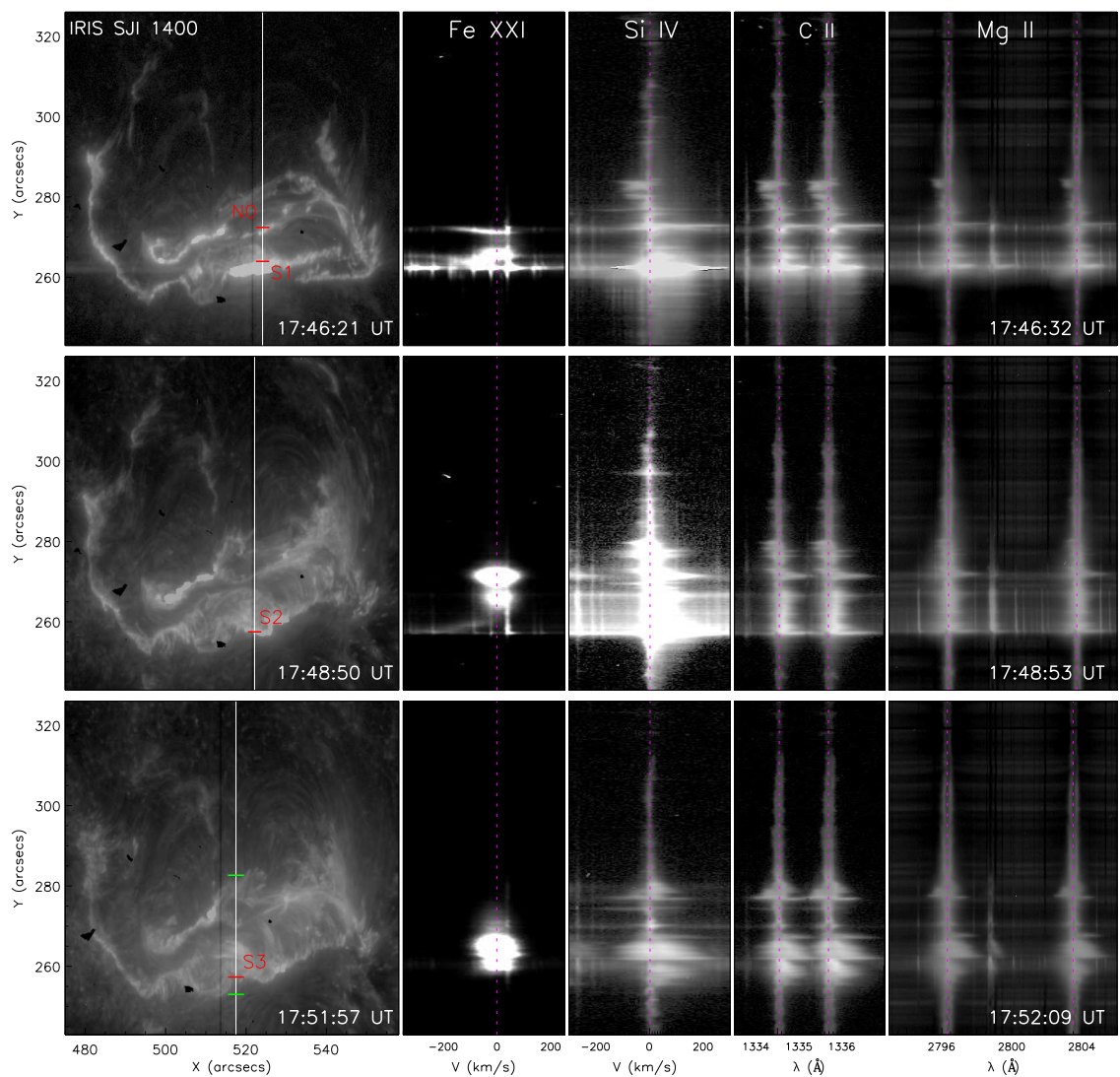}
\caption{IRIS slit-jaw 1400 \AA~images (column 1) and the slit spectra of Fe {\sc xxi}, Si {\sc iv}, C {\sc ii}, and Mg {\sc ii} h\&k lines (columns 2-5) in the rise phase, at the peak time, and in the decay phase of the flare, respectively. In the slit-jaw images, the white lines indicate the slit positions, while the red bars show four ribbon pixels whose line profiles are plotted in Figures \ref{profile_n0}--\ref{profile_s3}, respectively. The two green bars show the spatial range along the slit where the Doppler velocity and line width are presented in Figure \ref{iris_map}. In the spectrum panels, the pink dotted lines mark the reference wavelength position of each line.}
\label{iris_spec}
\end{center}
\end{figure*}

\begin{figure*}[htb]
\begin{center}
\includegraphics[width=16cm]{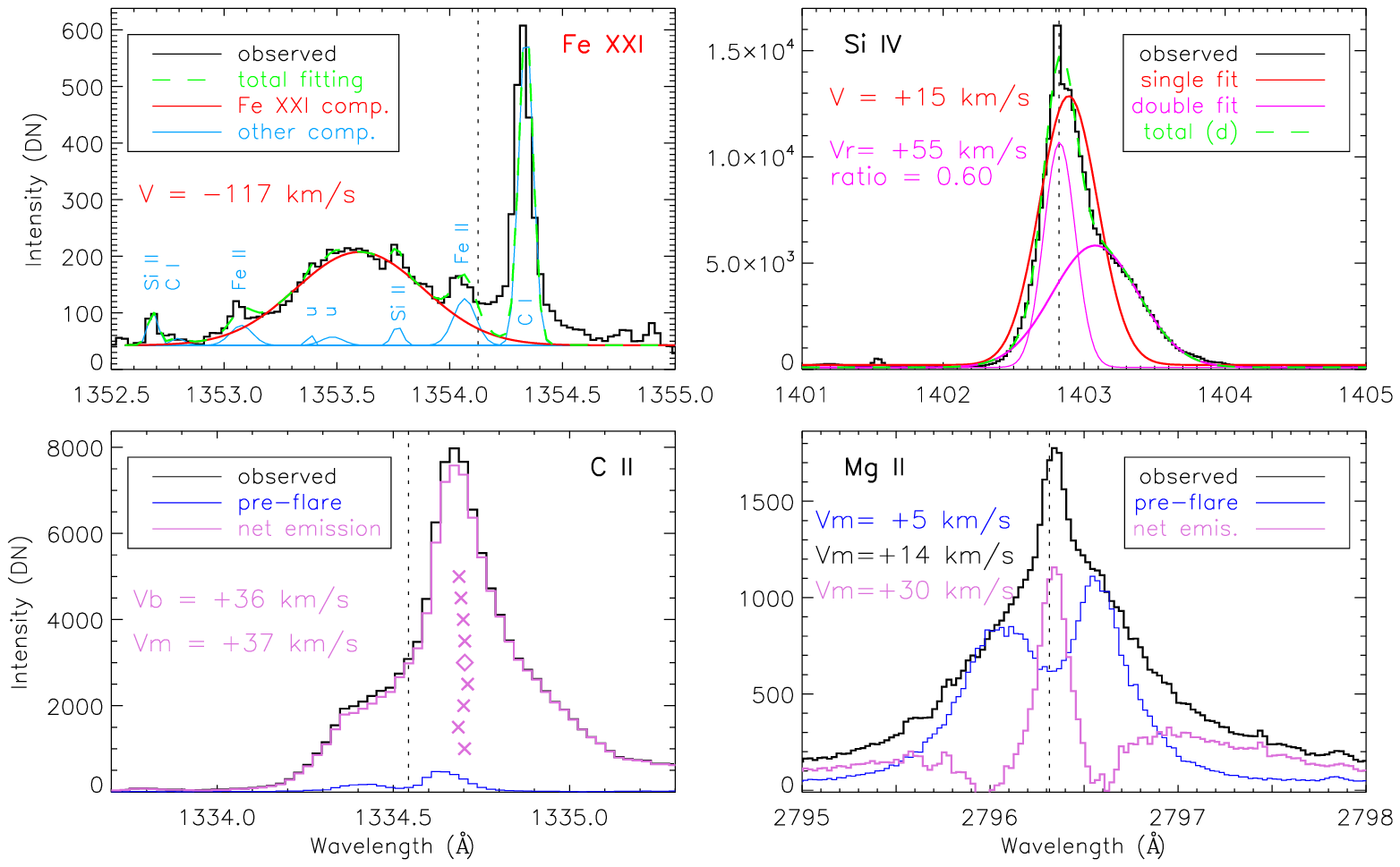}
\caption{Line profiles of Fe {\sc xxi}, Si {\sc iv}, C {\sc ii}, and Mg {\sc ii} k at ribbon pixel N0 marked in Figure \ref{iris_spec}. The black curves refer to observed line profiles, and colored curves to fitted line profiles or flare-net/pre-flare emissions. The vertical dotted lines indicate the reference wavelength for each line. The Doppler velocities ($V$ or $V_r$) derived from Gaussian fitting are given in the Fe {\sc xxi} and Si {\sc iv} panels, and the velocities from bisector ($V_b$, at the $\sim$40\% maximum intensity marked by the diamond) or moment ($V_m$) method given in the C {\sc ii} and Mg {\sc ii} panels. Negative and positive velocity values represent blueshifts and redshifts, respectively.}
\label{profile_n0}
\end{center}
\end{figure*}

\begin{figure*}[htb]
\begin{center}
\includegraphics[width=16cm]{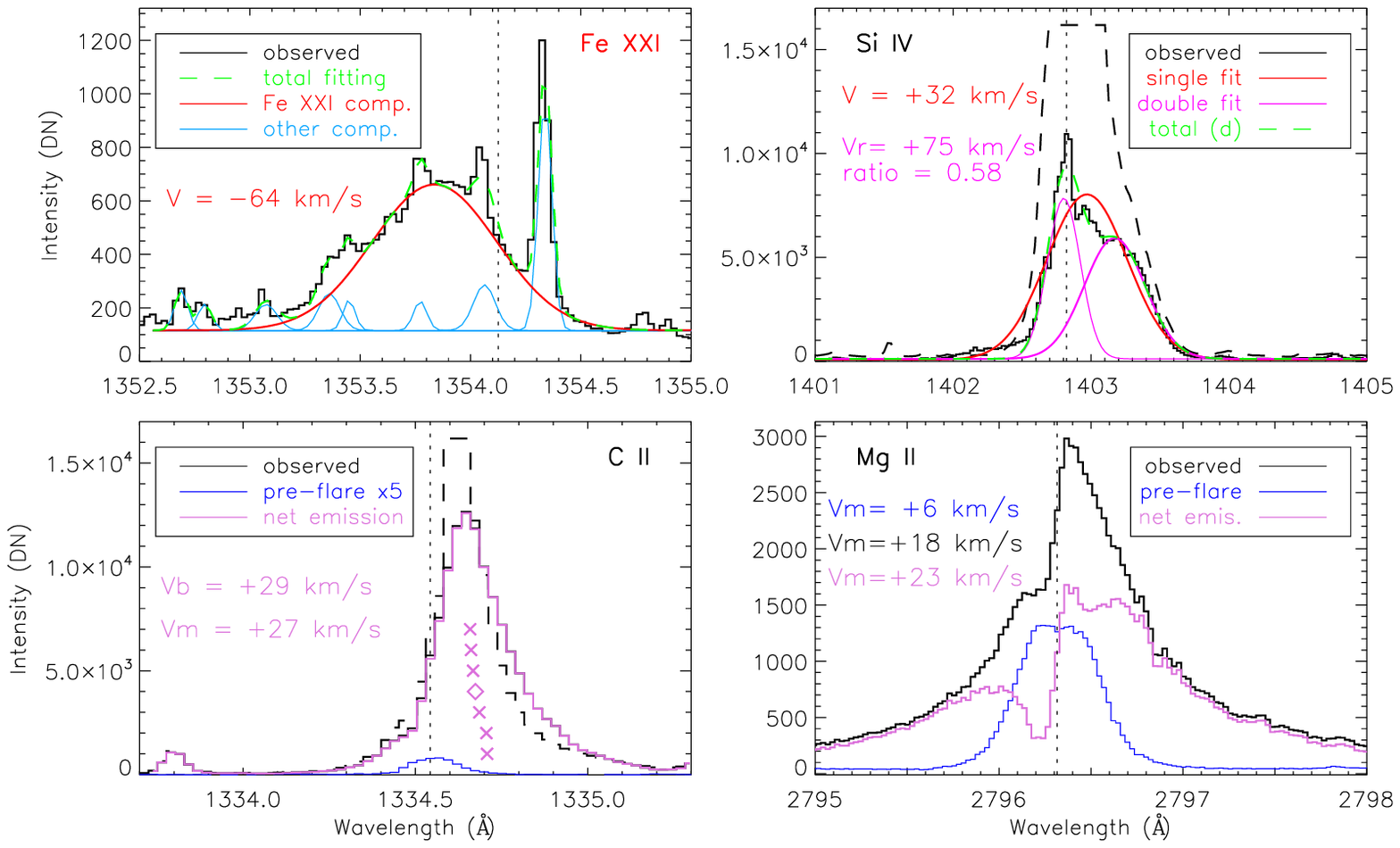}
\caption{Line profiles of Fe {\sc xxi}, Si {\sc iv}, C {\sc ii}, and Mg {\sc ii} k at ribbon pixel S1 marked in Figure \ref{iris_spec}. The black curves refer to observed line profiles, and colored curves to fitted line profiles or flare-net/pre-flare emissions. Note that there shows a saturation in the cores of the Si {\sc iv} and C {\sc ii} lines (black dashed curves), and we analyze the unsaturated profiles closest to S1 instead. The vertical dotted lines indicate the reference wavelength for each line. The Doppler velocities ($V$ or $V_r$) derived from Gaussian fitting are given in the Fe {\sc xxi} and Si {\sc iv} panels, and the velocities from bisector ($V_b$, at the $\sim$40\% maximum intensity marked by the diamond) or moment ($V_m$) method given in the C {\sc ii} and Mg {\sc ii} panels. Negative and positive velocity values represent blueshifts and redshifts, respectively.}
\label{profile_s1}
\end{center}
\end{figure*}

\begin{figure*}[htb]
\begin{center}
\includegraphics[width=16cm]{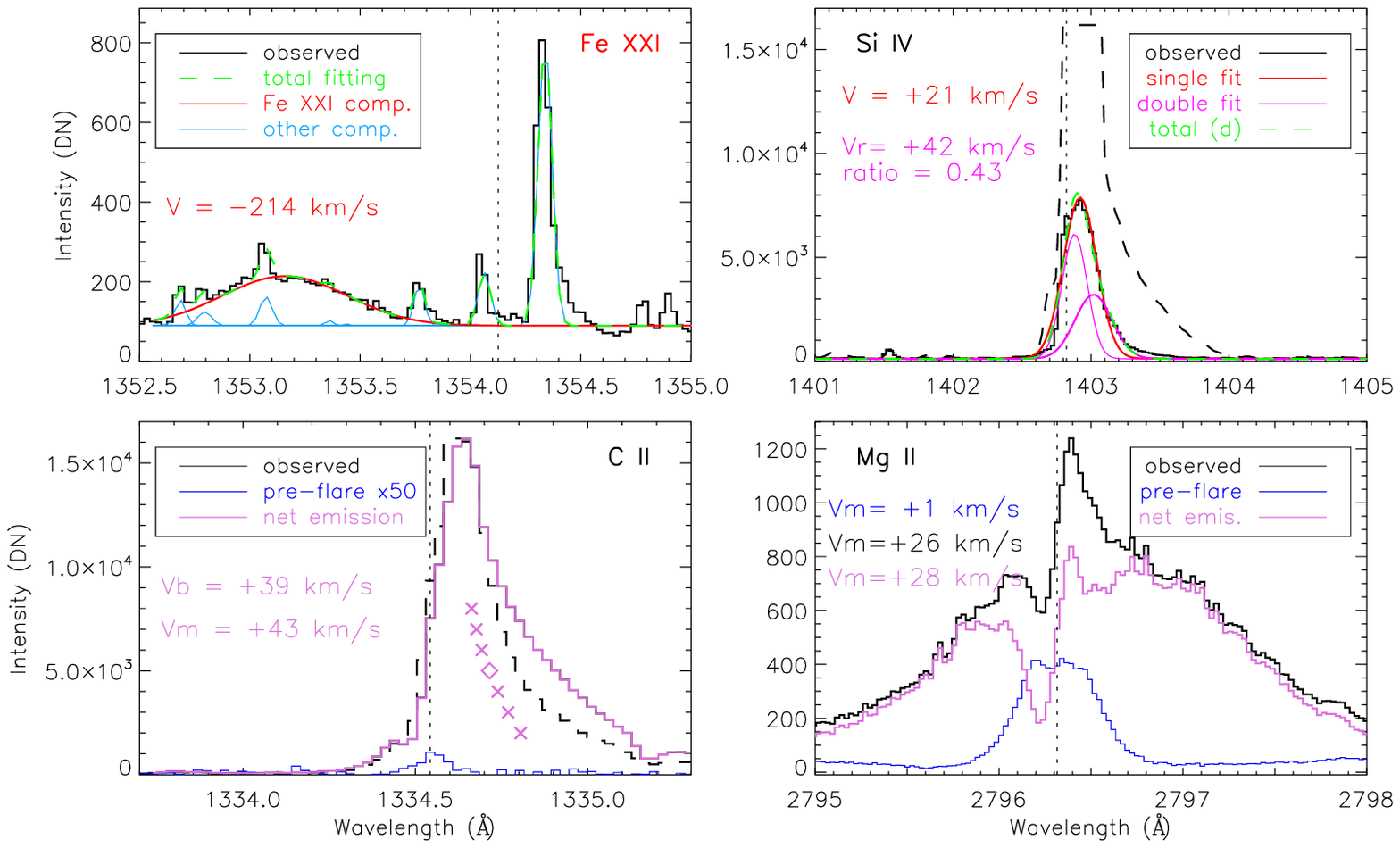}
\caption{Line profiles of Fe {\sc xxi}, Si {\sc iv}, C {\sc ii}, and Mg {\sc ii} k at ribbon pixel S2 marked in Figure \ref{iris_spec}. The black curves refer to observed line profiles, and colored curves to fitted line profiles or flare-net/pre-flare emissions. Note that there shows a saturation in the cores of the Si {\sc iv} and C {\sc ii} lines (black dashed curves), and we analyze the unsaturated profiles closest to S2 instead. The vertical dotted lines indicate the reference wavelength for each line. The Doppler velocities ($V$ or $V_r$) derived from Gaussian fitting are given in the Fe {\sc xxi} and Si {\sc iv} panels, and the velocities from bisector ($V_b$, at the $\sim$40\% maximum intensity marked by the diamond) or moment ($V_m$) method given in the C {\sc ii} and Mg {\sc ii} panels. Negative and positive velocity values represent blueshifts and redshifts, respectively.}
\label{profile_s2}
\end{center}
\end{figure*}

\begin{figure*}[htb]
\begin{center}
\includegraphics[width=16cm]{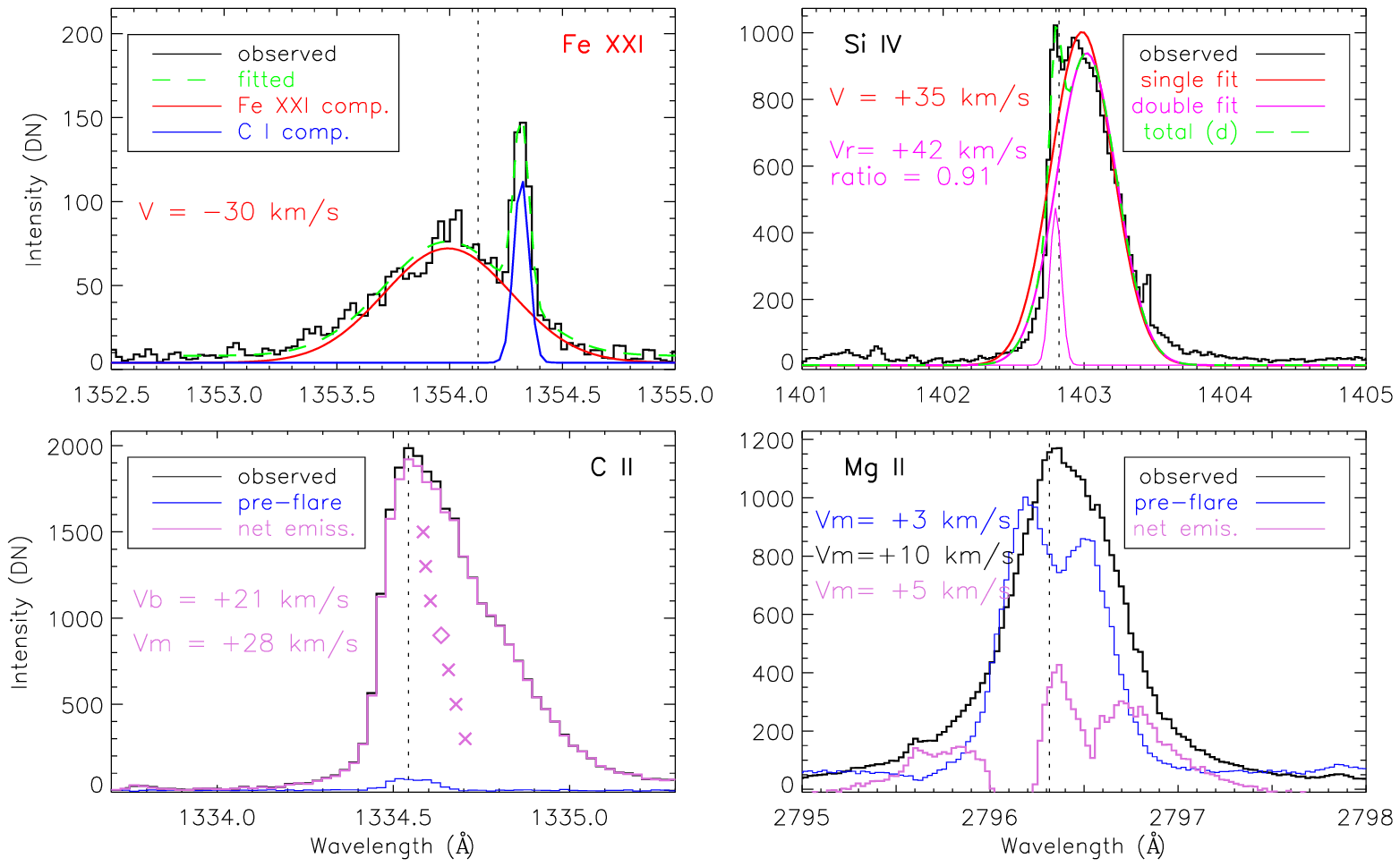}
\caption{Line profiles of Fe {\sc xxi}, Si {\sc iv}, C {\sc ii}, and Mg {\sc ii} k at ribbon pixel S3 marked in Figure \ref{iris_spec}. The black curves refer to observed line profiles, and colored curves to fitted line profiles or flare-net/pre-flare emissions. The vertical dotted lines indicate the reference wavelength for each line. The Doppler velocities ($V$ or $V_r$) derived from Gaussian fitting are given in the Fe {\sc xxi} and Si {\sc iv} panels, and the velocities from bisector ($V_b$, at the $\sim$40\% maximum intensity marked by the diamond) or moment ($V_m$) method given in the C {\sc ii} and Mg {\sc ii} panels. Negative and positive velocity values represent blueshifts and redshifts, respectively.}
\label{profile_s3}
\end{center}
\end{figure*}

\begin{figure*}[htb]
\begin{center}
\includegraphics[width=16cm]{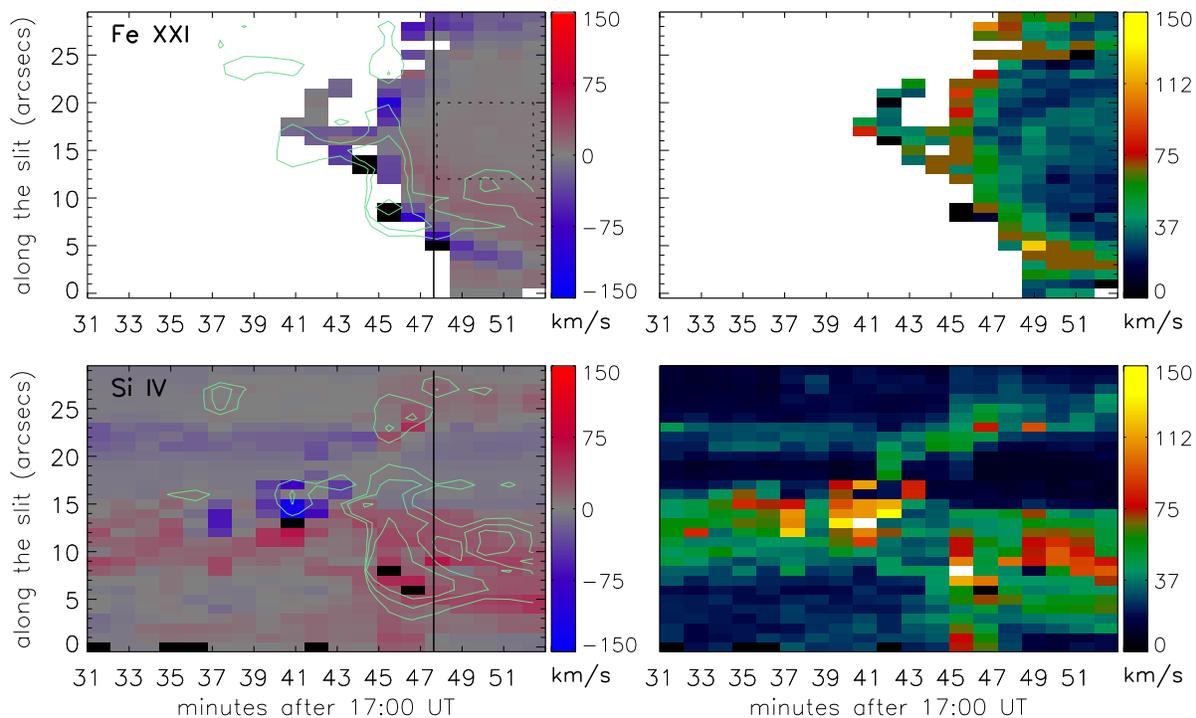}
\caption{Evolution of Doppler (left columns) and non-thermal (right columns) velocities in the Fe {\sc xxi} and Si {\sc iv} lines along the slit indicated by the two green bars in Figure \ref{iris_spec}. Overplotted on the Doppler velocity map in the Fe {\sc xxi} line are the intensity contours of the Mg {\sc ii} h line, and those on the Si {\sc iv} Doppler velocity map are the intensity contours of the Si {\sc iv} line itself that is slightly saturated. The vertical solid line represents the {\em GOES} soft X-ray peak time. In the Doppler velocity panel of Fe {\sc xxi}, the black data points indicate those Doppler blueshift velocities beyond the colorbar range. The box with dotted line shows the loop-top region that is used to obtain the average line center as the reference wavelength of Fe {\sc xxi}.}
\label{iris_map}
\end{center}
\end{figure*}

\begin{figure*}[htb]
\begin{center}
\includegraphics[width=16cm]{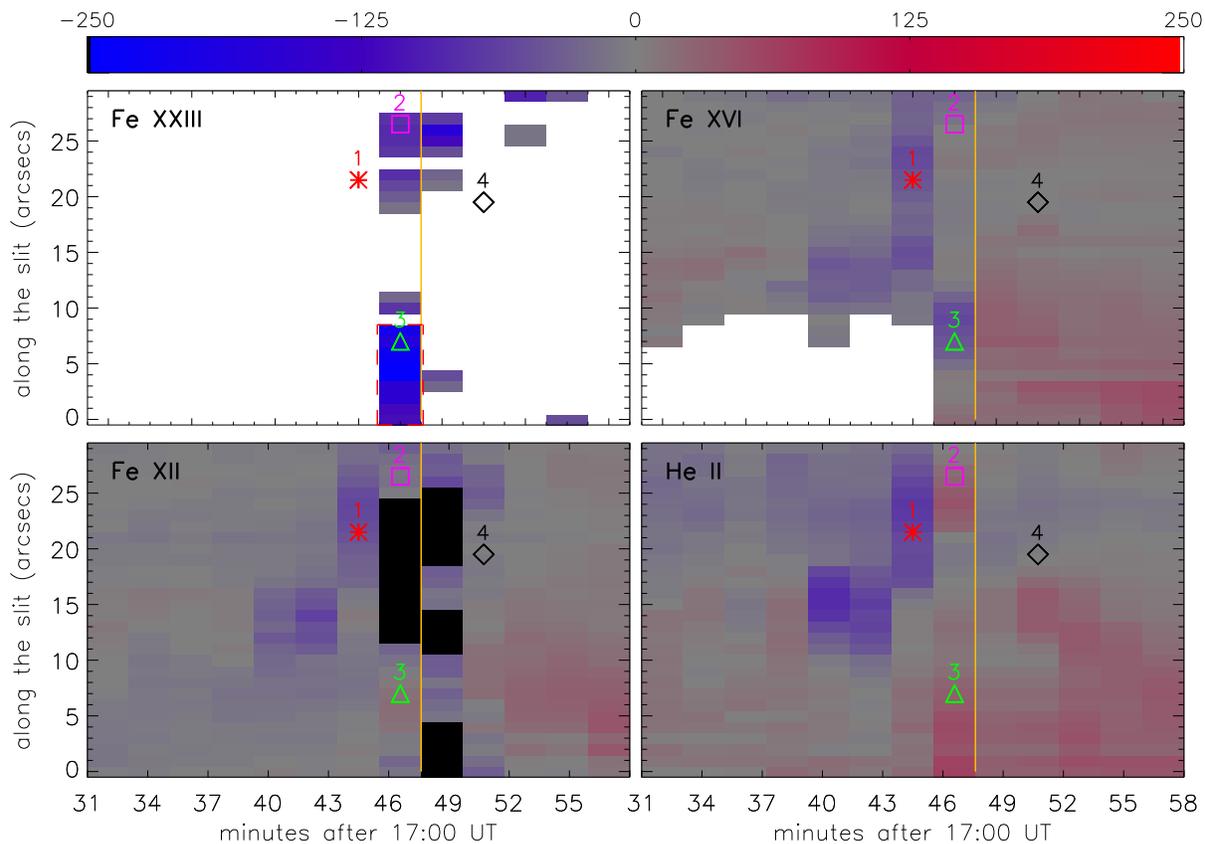}
\caption{Evolution of the Doppler velocity along the slit in four EIS lines. The slit position is around the same as the IRIS slit position shown in Figure \ref{iris_map}. The yellow line marks the {\em GOES} soft X-ray peak time. The four labeled locations refer to places where the Doppler velocities are shown in Figure \ref{eis_iris}. Note that, the Fe {\sc xxiii} velocities represent the shifts of blueshifted components, which are derived from a double Gaussian fitting. The pixels within the red rectangular box in the Fe {\sc xxiii} panel show double-peak line profiles (see the top left panel in Figure \ref{eis_prof}), and the other pixels only show a line profile with a weak blue wing enhancement (see the top middle panel in Figure \ref{eis_prof}).}
\label{eis_tmap}
\end{center}
\end{figure*}

\begin{figure*}[htb]
\begin{center}
\includegraphics[width=16cm]{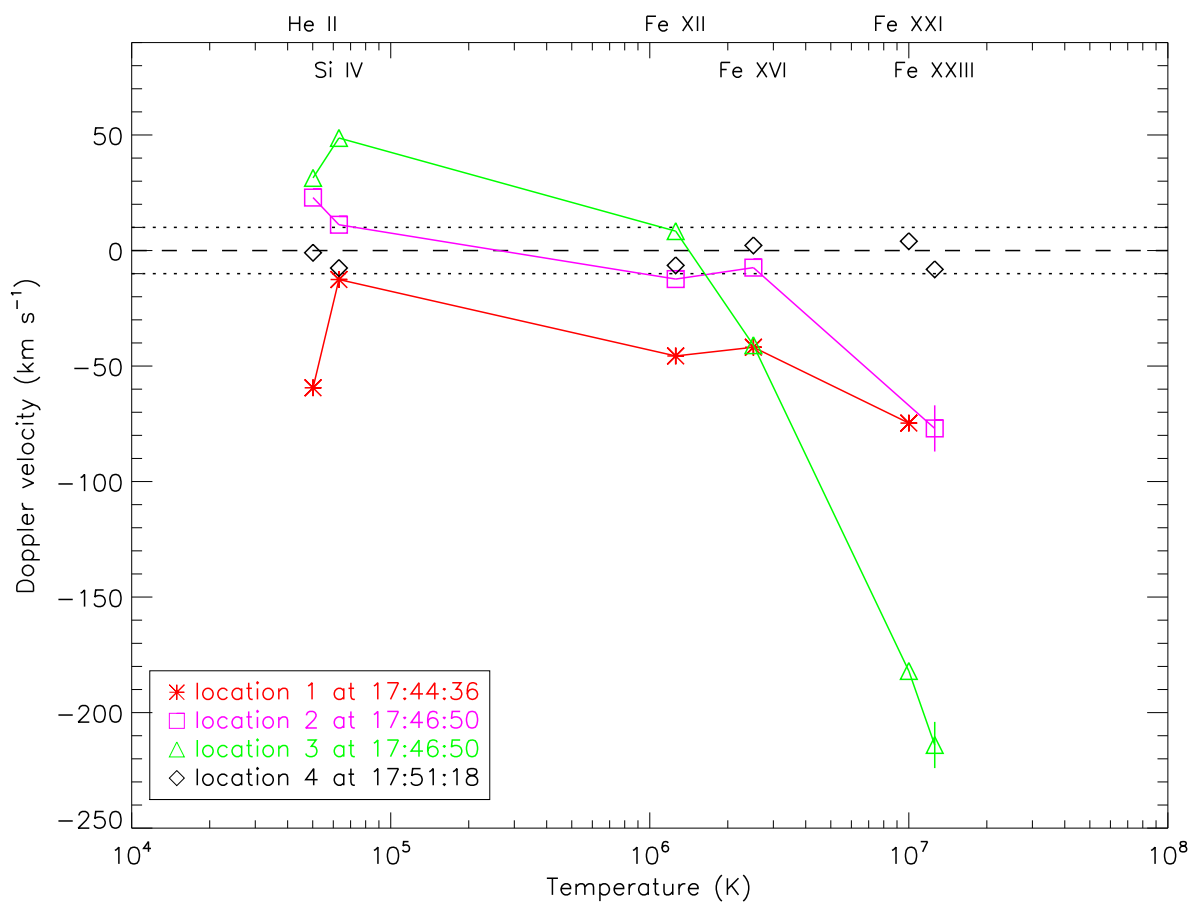}
\caption{Doppler velocity as a function of temperature at four locations (marked in Figure \ref{eis_tmap}), derived from EIS EUV and IRIS UV lines. The dashed line refers to the zero velocity, and the two dotted lines show the range of velocity uncertainties. Negative velocity values represent blueshifts. Note that all the velocities are derived from a single Gaussian fitting, except the two data points marked by a vertical bar in the Fe {\sc xxiii} line at locations 2 and 3, which are derived from a double Gaussian fitting.}
\label{eis_iris}
\end{center}
\end{figure*}

The Mg {\sc ii} and C {\sc ii} line profiles become pure emission in the core from a central reversal during the flare, which may suggest a greatly reduced opacity \citep{tian14a}. The opacity reduction might be caused by energy deposition in the line formation layer. Furthermore, a sharp change in the spectra in these lines (see Figure \ref{iris_spec}) may indicate mass motions with a very large velocity gradients, like shocks. During the flare, these lines exhibit broadened wings and red asymmetry, in addition to the line center shift. In addition, the Mg {\sc ii} subordinate lines are also significantly enhanced, providing more diagnostic power of the flaring atmosphere. IRIS shows a great potential to diagnose and help understand heating and dynamics of the lower atmosphere during solar flares.

In the present study, we only focus on the spectra in the UV and EUV bands from two spectrographs, IRIS and EIS. In fact, for this X1.0 flare, there are also joint observations from some other instruments, such as the Solar Optical Telescope and the Dunn Solar Telescope, as well as AIA and the Helioseismic and Magnetic Imager. All these instruments provide rich data for us to study the different aspects of solar flares, which will be performed in a future work.


\acknowledgments
The authors thank the anonymous referee for constructive comments to improve the manuscript. Y.L. thanks Hui Tian and Sarah Jaeggli for the IRIS data processing and Dana Longcope for scientific discussions. Y.L. also thanks Peter Young for his valuable comments to improve the manuscript. This project was supported by NSFC under grants 10933003, 11373023, 11103008, and 11403011, and by NKBRSF under grants 2011CB811402 and 2014CB744203. Y.L. is also supported by the Postdoctoral Science Foundations from Jiangsu Province and China Postdoctoral Office, by the Fundamental Research Funds for the Central Universities, and by the Office of China Postdoctoral Council (No. 29) under the International Postdoctoral Exchange Fellowship Program 2014. IRIS is a NASA small explorer mission developed and operated by LMSAL with mission operations executed at NASA Ames Research center and major contributions to downlink communications funded by the Norwegian Space Center (NSC, Norway) through an ESA PRODEX contract. \textit{Hinode} is a Japanese mission developed and launched by ISAS/JAXA, collaborating with NAOJ as a domestic partner, and NASA (USA) and STFC (UK) as international partners. \textit{SDO} is a mission of NASA's Living With a Star Program.

\bibliographystyle{apj}

\begin{table}[htb]
\begin{center}
\caption{Spectral Lines Used in This Study \citep{depo14,youn07}}
\label{line}
\begin{tabular}{lccc}
\tableline
\tableline
\multicolumn{1}{l}{Ion}  &  Wavelength (\AA)  &  log $T_{max}$~(K)  &  Instrument \\
\tableline
Mg {\sc ii} h     & 2803.5   & 4.0 & IRIS \\
Mg {\sc ii} k     & 2796.4   & 4.0 & IRIS \\
C {\sc ii}          & 1334.5   & 4.3 & IRIS \\
C {\sc ii}          & 1335.7   & 4.3 & IRIS \\
He {\sc ii}        & 256.32	& 4.7 & EIS \\
Si {\sc iv}        & 1402.8    & 4.8 & IRIS \\
Fe {\sc xii}      & 192.39	& 6.1 & EIS \\
Fe {\sc xvi}     & 262.98	& 6.4 & EIS \\
Fe {\sc xxi}     & 1354.1   & 7.0 & IRIS \\
Fe {\sc xxiii}    & 263.76	& 7.1 & EIS \\
\tableline
\end{tabular}
\end{center}
\end{table}

\end{document}